\documentclass[10pt]{article}
\usepackage{graphicx}

\begin{document}

\title{\vspace{-2.0cm} Numerical Model for the Deformation of Nucleated Cells by Optical Stretchers}
\date{}
\maketitle
\vspace{-15mm}

\begin{center}

\author{Ihab Sraj \\ Division of Physical Sciences and Engineering \\ 
King Abdullah University of Science and Technology 
\\ Thuwal, Saudi Arabia \\\textit{ihab.sraj@gmail.com}}\\
\vspace{5mm}
\author{Joshua Francois \\ Department of Mechanical Engineering \\ 
University of Maryland Baltimore County \\ Baltimore, Maryland 21250, USA}\\ \textit{fjoshua1@umbc.edu}\\
\vspace{5mm}
\author{David W.M. Marr \\ Department of Chemical and Biological Engineering \\
 Colorado School of Mines, \\ Golden, Colorado 80401, USA}\\\textit{dmarr@mines.edu} \\
\vspace{5mm}
\author{Charles D. Eggleton \\ Department of Mechanical Engineering \\ 
University of Maryland Baltimore County 
\\ Baltimore, Maryland 21250, USA \\ \textit{eggleton@umbc.edu}}
\end{center}



\begin{abstract}
In this paper, we seek to model the deformation of nucleated
cells by single diode-laser bar optical stretchers. We employ a recently 
developed computational model, the Dynamic Ray-Tracing method, to 
determine the stress distribution induced by the applied optical forces
on a capsule encapsulating a nucleus of different optical properties. 
These forces are shape dependent and can deform real non-rigid objects; thus resulting 
in a dynamically changing optical stress distribution with cell and nucleus deformation.
Chinese hamster ovary cell is a common biological cell
that is of interest to the biomedical community because of their use in recombinant
protein therapeutics and is an example of a nucleated cell.
To this end, we model chinese hamster ovary cells 
as two three-dimensional elastic capsules of variable inner capsule 
size immersed in a fluid where the hydrodynamic 
forces are calculated using the Immersed Boundary Method. Our results show
that the presence of a nucleus has a major effect on the force distribution
on the cell surface and the net deformation. Scattering 
and gradient forces are reported for different nucleus sizes and the effect of 
nucleus size on the cell deformation is discussed. 
\end{abstract}

\section{Introduction}
The ability to trap particles using laser light was discovered by 
Arthur Ashkin in 1970~\cite{Ashkin1970}. In this, gradient forces are 
created at the surface of transparent particles suspended in a medium of different 
refractive index and situated within a light gradient. In the ray-optics regime, 
where the particles size is much larger than the light wavelength~\cite{Hulst1957},
refraction of light rays of different intensities at 
the surface and within the particles results in a change in 
the total momentum between the entering and exiting light beam. These 
gradient forces on the order of picoNewtons
are capable of drawing microscopic particles into 
a region of highest light intensity~\cite{Ashkin1986}. 
Scattering forces are also created that accelerate the particle 
in the direction of beam propagation towards its focus~\cite{Ashkin1971}. 
With these opposing mechanisms, an equilibrium position is reached and the particle is held fixed (trapped)
in the center of the beam focus as the light rays passing
through and exiting the particle exert forces that are balanced 
with no net change in momentum.

Optical traps or tweezers used to manipulate microscopic objects without 
any mechanical contact have become a major tool in biological research over 
the last thirty years. Cells have been stretched~\cite{Guck2001}, folded~\cite{Gu2007} 
and even rotated \cite{Gu2007,Mohanty2004} using single and multiple optical tweezers. 
Recently, the technique has been extended to study the properties of cells
by observing their deformation~\cite{Bronkhorst1995,Guck2005,Sraj2010a}.
Guck \emph{et al.} developed an optical stretcher that uses two counterpropagating
diverging laser beams to trap cells individually along the aligned laser
beams axis~\cite{Guck2001}. This optical stretcher has also been used to deform cells 
and to measure their membrane properties using a simple numerical model~\cite{Guck2005}. 
Extending this, Sraj \emph{et al.} developed a high-throughput optical stretcher using 
a single linear diode bar to trap and stretch bovine red blood cells (RBC)~\cite{Sraj2010a}.

Theoretical and numerical studies have also been conducted to determine the 
induced optical force distribution. Using analytical solutions of the governing 
optical equations, Ashkin was the first to perform
calculation of the total forces of a single-beam gradient laser on solid spheres in the 
ray-optics (RO) regime~\cite{Ashkin1992}. Guck \emph{et al.} determined the local optical force distribution 
by the dual optical stretcher on spherical cells 
using the RO technique before cells begin to deform \cite{Guck2001} to determine the stiffness of RBCs~\cite{Guck2005}. 
In their method, the effect of subsequent deformation on the calculation of force 
distribution was neglected and constant rigid spherical cell morphology was assumed due to limitations 
in the method. Deformability of biological cells can result in a shape
change under the influence of external flows or applied forces and thus the local force
distribution and total trapping forces can change significantly with cell deformation~\cite{Sraj2010b}. 
To take this into account, the RO method has been improved to include different cell shapes such as oblate spheroids
\cite{Sosa-Martinez2009} and even cylinders \cite{Gauthier1997a}. As an analytical method, RO
remains a difficult approach for calculating the forces on more complex cell shapes like the RBC bi-concave discoid shape and deformable cells. 

To overcome these issues, we recently developed and implemented
a dynamic ray-tracing (DRT) approach~\cite{Sraj2010b} that, in addition to finding transient optical 
forces on deformable cells, solves for fluid-cell interactions \cite{Peskin1989,Eggleton1998}.
DRT offers the possibility of simulating different phenomena occurring in optical systems
such as erythrocyte deformation in high-throughput optical stretchers~\cite{Sraj2012} and optical levitation~\cite{Chang2012}. 
The approach allows one to assess cell deformability and to investigate the optical parameters to better design traps and manipulate cells prior to
performing experiments. In addition, due the vector-based nature, DRT allows the calculations of 
both anisotropic and inhomogeneous structures, cases that exist in real systems.

CHO cells are the most commonly used mammalian host 
for industrial production of recombinant protein therapeutics \cite{Karthik2007}
and have been used in related genetics studies~\cite{Tjio1958}. Because of their importance, a number of 
previous studies have investigated the optical forces on CHO cells. For example, Wei \emph{et al.} used a 
fiber-optic dual-beam trap to capture chinese hamster ovary (CHO) cells
and determine the associated three-dimensional optical force field \cite{Wei2006}.
Chang \emph{et al.} developed a model 
based on RO to calculate the optical force upon a solid spherically-symmetric 
multilayer sphere~\cite{Chang2006}. 
This study showed that the magnitude of optical forces are three times smaller 
than that upon a polystyrene bead of the same size and that the 
distribution of optical forces is much different 
from that upon a uniform particle. Recently, Kim \emph{et al.} 
computed the optical force on a pair of concentric spheres in a 
focused beam and determined the influence of refractive index differences and relative size between the inner and 
outer spheres on the optical force~\cite{Kim2012}.

All of these studies however did not take into account the deformability of both the cell
and its encapsulated nucleus. We show here how the presence 
of a nucleus inside deformable cells leads to alteration in the propagation 
of light rays due to the additional internal surface 
and the additional medium 
of different refractive index~\cite{Meyer1975}. Here, the variation in the nucleus size
may significantly influence the optical forces and ultimately the net deformation of both the cell and the nucleus itself.
\section{Numerical method}

Simulating deformation of a cell via optical stretchers requires 
a two-step method to first determine optical stresses induced by 
the interaction of light with the cell surface and then model the 
cell-fluid interactions. Because the distribution of optical stresses 
is dependant on the shape of the cell that in turn changes during 
deformation, these two steps are done alternately until a steady 
state shape is reached. In the case of nucleated cells, optical
stress calculation is additionally challenging due to the different external
and internal morphologies in addition to membrane deformability. 
For this purpose, we resort to DRT to 
determine the optical stresses induced on the surfaces of deformable 
nucleated cells by optical stretchers~\cite{Sraj2010a,Sraj2012}. DRT,
unlike the traditional RO method, is vector-based and is 
capable of determining the optical forces on cells of arbitrary shape 
and morphology. Cell-fluid interaction and the hydrodynamics, on the
other hand, are solved using the Immersed Boundary Method (IBM). 
In this section we briefly describe the two methods.

\subsection{Dynamic Ray Tracing}
\label{sec:drt}

DRT is a vector-based method developed by Sraj 
\emph{et al.}~\cite{Sraj2010b,Sraj2012} to determine optical forces on 
the surface of any arbitrary shaped cell including deformable cells 
with asymmetrical geometries \cite{Chang2012}. Briefly, DRT 
considers a finite number of rays issued from a light source with given 
intensities and known direction. These rays are treated as vectors and
traced as they intersect a surface. A ray-triangle intersection algorithm is 
then employed to determine the location of intersection where the surface 
is divided into triangular elements for this purpose. Geometrical optics 
laws are then applied to find the refraction and reflection angle. 
Consequently the vectors of the rays are updated and the procedure is repeated
till each ray exits the cell. From the direction of the light rays
within the cell one can calculate the trapping efficiency $Q$, a 
dimensionless factor representing the amount of momentum transferred 
\cite{Guck2001,Sraj2010b}. The trapping efficiency $Q$ is independent of the 
laser power used and depends only on the object geometry and reflectance of 
the medium. 
Elemental optical forces at any location of the cell are therefore found 
regardless of the initial cell shape. Rays are traced at any surface where 
$Q$ is multiplied by a factor to account for energy loss from previous 
refractions. Internal and external reflections within the cell are neglected 
as their effects rapidly diminish. Optical forces can be expressed as

\begin{equation}
F _{optical} = \frac{n_{m}QP}{c},
\end{equation}	
where $n_m$ is the index of refraction of the buffer medium, $c$ the 
speed of light in vacuum, and $P$ the laser power. It is important to 
note that the resulting optical forces are added to the Navier-Stokes 
equations as body forces as described below.
 
This method has been validated and applied to cells of different initial shapes
\cite{Sraj2010b} and different optical applications \cite{Sraj2012,Chang2012}. 
Here, DRT is used to model nucleated cells by 
considering two concentric spheres of different sizes, one representing 
the cell surface and another representing the nucleus. DRT is then employed
to determine optical stresses induced on both surfaces. 
\subsection{Immersed Boundary Method}
\label{subsection:IBM}
The IBM is a cell-fluid interaction solver that has been used extensively to simulate 
biological systems such as cell adhesion \cite{Gupta2010}, cell adhesion in 
atomic force microscopy measurements \cite{Sraj2011} and red blood cell motion
through microvascular bifurcation \cite{Xiong2012}. IBM splits the numerical 
solution onto two grids: a stationary grid that has a fixed position with time 
representing the three-dimensional fluid domain and a moving grid representing the 
two-dimensional immersed boundary.

To this end, a cell is modeled as an elastic membrane that is 
deformable by any applied stress. The membrane is discretized 
into a finite number of flat triangular elements that remain 
flat after deformation. This approximation is valid given 
that the local radius of curvature during deformation is much 
larger than the membrane thickness and that bending stresses 
are negligible. Elastic forces at the discrete membrane nodes 
are found from their displacement (deformation) using a finite 
element model. We adopt an approach developed by Charrier 
\emph{et al.}~\cite{Charrier1989} and Shrivastava and Tang~\cite{Shrivastava1993}
that uses the principle of 
virtual work to find those forces from an appropriate strain 
energy density function. These forces and any external 
applied forces such as the optical forces are then 
distributed onto the fluid grid using an appropriate discrete 
delta function and added to the Navier-Stokes equations as 
body forces. The discrete delta function ensures that only 
membrane nodes in the sphere of influence of the fluid grid 
make a contribution to the local body forces. The Navier-Stokes 
equations are then solved for the fluid velocity.

The no-slip boundary condition at the membrane surface is 
satisfied by allowing the membrane nodes to move with the local 
fluid velocity. The velocity of the membrane is found by summing 
of the velocities at the fluid grid nodes weighted by the same 
discrete delta function used for the distribution of body forces. 
This again ensures that only fluid grid nodes in the sphere of 
influence of the membrane node make a contribution to its velocity. 
Membrane nodes are then moved with the calculated velocity for 
one time step to a new position giving a new membrane shape. 
The procedure is repeated and elastic forces and optical forces 
are then calculated as described above to advance the flow for 
another time step.

\section{Model parameters}

The CHO cell is a typical example
of a nucleated cell. The size of such cells
vary with radius $r_{cell}$ ranging from $5-7.5~\mu m$~\cite{Han2006}
and nucleus radius $r_{nuc}$ varying following the
relationship~\cite{Brunsting1974}: 
\begin{equation}
r_{cell} = (1.38\pm 0.02)r_{nuc} + (0.03\pm 0.05). 
\label{eq:ratio}
\end{equation}
The refractive index of the cytoplasm has been measured and reported
as $n_{cyt}=1.37$ while that of the nucleus has been found to be
slightly greater $n_{nuc} = 1.392$~\cite{Brunsting1974}. As the 
size of the cell has an effect on the forces induced 
by optical stretchers (more surface area intuitively results in large forces)
and hence on the cell deformation, we seek to investigate the 
impact of the presence of the nucleus and its size on the optical 
stress distribution and resulting cellular deformation. For this 
purpose, we model CHO cells as two three-dimensional (3D) concentric elastic spherical 
capsules. The radius of the outer capsule is fixed and taken as the CHO cell 
average radius $r_{cell} = 5.6~\mu m$; however, the radius of the inner 
capsule representing the nucleus is varied from 
$r_{nuc} = 1.1-5~\mu m$. The ratio of the radius of the
cell to the radius of the nucleus is denoted by 
$r = \frac{r_{nuc}}{r_{cell}}$. We note that a typical CHO cell 
has radii ratio of $r = 0.72$ from Equation~\ref{eq:ratio}.

\begin{figure}[htbp]
\centering 
\includegraphics[width=100mm]{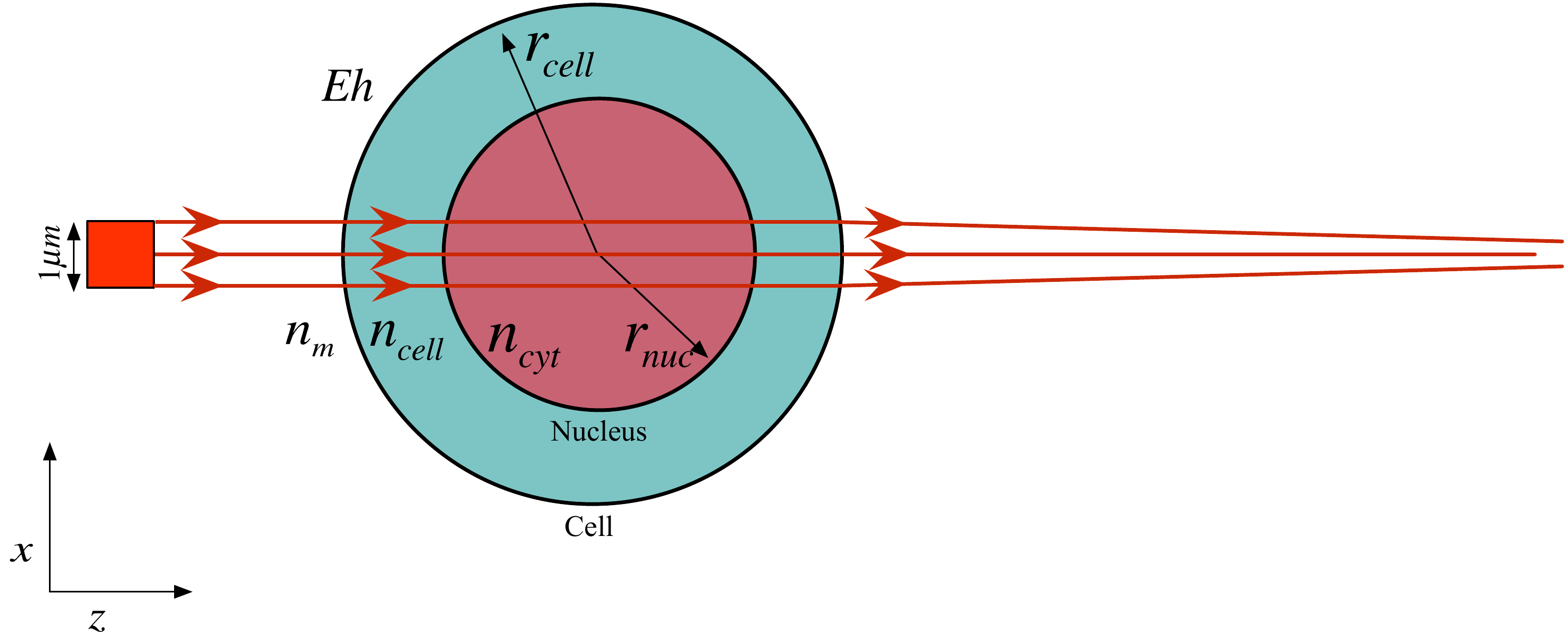}
\caption{CHO cell model: $n_m<n_{cell}<n_{cyt}$ in a linear diode stretcher.} 
\label{fig:model} 
\end{figure}

In our calculations, cells are assumed initially
trapped and situated at the center of a laser beam 
created with a single linear diode bar of wavelength $\lambda = 808~nm$
and power $P = 12.5~ mW/\mu m$. The length of the diode lies in the 
$y$-axis and the laser beam direction is along the $z$-axis as shown 
in Fig.~\ref{fig:model}. The cell is assumed to be immersed in an aqueous medium of 
refractive index $n_m=1.335$ that is lower than the refractive index of both 
the cell and its nucleus ($n_m<n_{cyt}<n_{nuc}$) (Fig.~ \ref{fig:model}). 

The hydrodynamics and cell mechanics are calculated using the IBM. Both fluids inside 
and outside the cell are assumed incompressible and Newtonian with identical density $\rho = 1~g/cm^3$ and viscosity $\mu = 0.8~cP$. The cell membrane 
is assumed of Neo-Hookean material, as appropriate for most biological cells, and can be 
characterized using solely its stiffness $Eh$. Unless otherwise noted, membrane 
stiffness is taken as $Eh = 0.1~dyn/cm$.  
For the purpose of quantifying cell deformation from optical or hydrodynamic forces, we use the Taylor 
deformation parameter defined as: $DF = (L-B)/(L+B)$, where $L$ and $B$ are the major 
and minor semi-axis of a capsule in the $x-z$ plane. When viscous 
stresses, elastic forces and optical forces are balanced, cells adopt a steady state 
shape denoted by $DF_{\infty}$. The uniform grid used for the fluid solver has $64^3$ 
nodes with a grid spacing of $r_{cell}/8$ while the finite element cell grid has $20482$ 
triangular elements. A time step of $10^{-5} s$ was used in all computations to ensure numerical stability.

\section{Results and discussions}

\subsection{Impact of nucleus size on optical forces}

As a first step, we investigate the effect of nucleus 
size on the optical forces initially induced at the cell 
surface. For reference, we employ DRT to determine the 
forces induced on a cell with no nucleus.
In this case, light rays emerging from the diode laser 
bar hit the front surface of the cell to create optical 
scattering forces along the laser beam axis whose 
direction is opposite to their propagation direction i.e. 
the negative $z-$axis direction as shown in 
Fig.~\ref{fig:model}. This is due to the momentum 
gained by the cell when the rays transit from a medium 
of lower refractive index to a medium of higher refractive index
(the cell cytoplasm $n_{cyt} = 1.37$)
~\cite{Guck2001}. Gradient optical forces are also created
with a sum equal to zero as the 
center of the cell is aligned with the center of the laser 
beam. After refraction, the rays continue to hit the 
back surface of the cell where they refract again and 
transfer momentum inducing scattering forces in the 
positive direction. The magnitude of the net scattering force at the 
back surface of the cell is, however, greater than the 
net scattering force at the front surface resulting in a net 
total scattering force in the positive $z-axis$ direction.
With our chosen laser, cell, and fluid properties, the 
magnitude of scattering force applied on the 
front surface is $23.1~pN$ while on the back surface is $26.7~pN$ with
a net scattering force of 
$3.6~pN$. The scattering forces would both stretch and translate the cell away 
from the light source. The net gradient forces are again
equal to zero on both the front and back surface of the cell; 
however, if we consider the net gradient force in the perpendicular
direction to the laser beam on one half of the cell we find it
equal to $20.7~pN$. The gradient forces contribute to the stretching of the cell
as well but have no translation effect.

In cells with a nucleus, light rays can hit up to four surfaces as 
shown in Fig.~\ref{fig:model} before exiting the cell from the back 
surface. This leads to scattering and gradient optical forces 
at both the outer cell and the nucleus surface. At the front cell surface, 
scattering forces remain unchanged and independent of the nucleus size as the rays first 
enter the cell as already described. As the refractive index of the nucleus is 
higher than that of the cytoplasm, scattering forces at the front surface are 
also in the negative direction while the same forces at the back surface are 
in the positive direction. However, the magnitude of the net scattering force on the 
nucleus is less than the net scattering force on the cell due to the smaller 
refractive index contrast between the nucleus and the cell 
compared with the cell and suspending medium ($\frac{n_{nuc}}{n_{cyt}}=\frac{1.392}{1.37} = 1.016$ versus 
$\frac{n_{cyt}}{n_m} = \frac{1.37}{1.335} = 1.030$ ). The magnitude of 
these forces depends on the nucleus size. For instance, 
a nucleus of radius $r_{nuc} = 4~\mu m$ experiences a scattering force of 
$4.54~pN$ on the front surface of the nucleus and a force of $5.24~pN$ on the back surface with a net scattering force on the 
nucleus of $0.7~pN$. When the nucleus size is increased to 
$r_{nuc} = 5~\mu m$, the net force on the nucleus increases 
to $8.75~pN$ and $10.4~pN$ on the front and back 
surface respectively with a total net force of $1.6~pN$. 
This increase in force has a significant effect on
the nucleus deformation and the 
forces induced on the cell itself as discussed in the following section. 

As the rays reach the cell back surface, the forces induced are influenced 
by the previous two refractions at the nucleus. As the nucleus size increases, 
optical forces on the nucleus increase, rays gain 
more energy, and we see subsequent increase in the magnitude of optical 
forces at the back cell surface. For a nucleus of radius $r_{nuc} = 4~ \mu m$ 
the net force at the back of the cell is $27.1~pN$ and $27.4~pN$ for $r_{nuc} = 5~ \mu m$. The maximum increase in the
net optical forces on the cell is thus $20.4\%$ compared with the nucleus-free case. 
Finally, for $r_{nuc} = 5~ \mu m$ the gradient forces calculated
on the cell slightly decreased to $20.2~pN$ due to the presence of the nucleus
while the gradient forces calculated on the nucleus increased to $8.6~pN$.

\begin{figure}[htbp]
\centering 
\begin{tabular}{cc}

\includegraphics[width=0.45\textwidth]{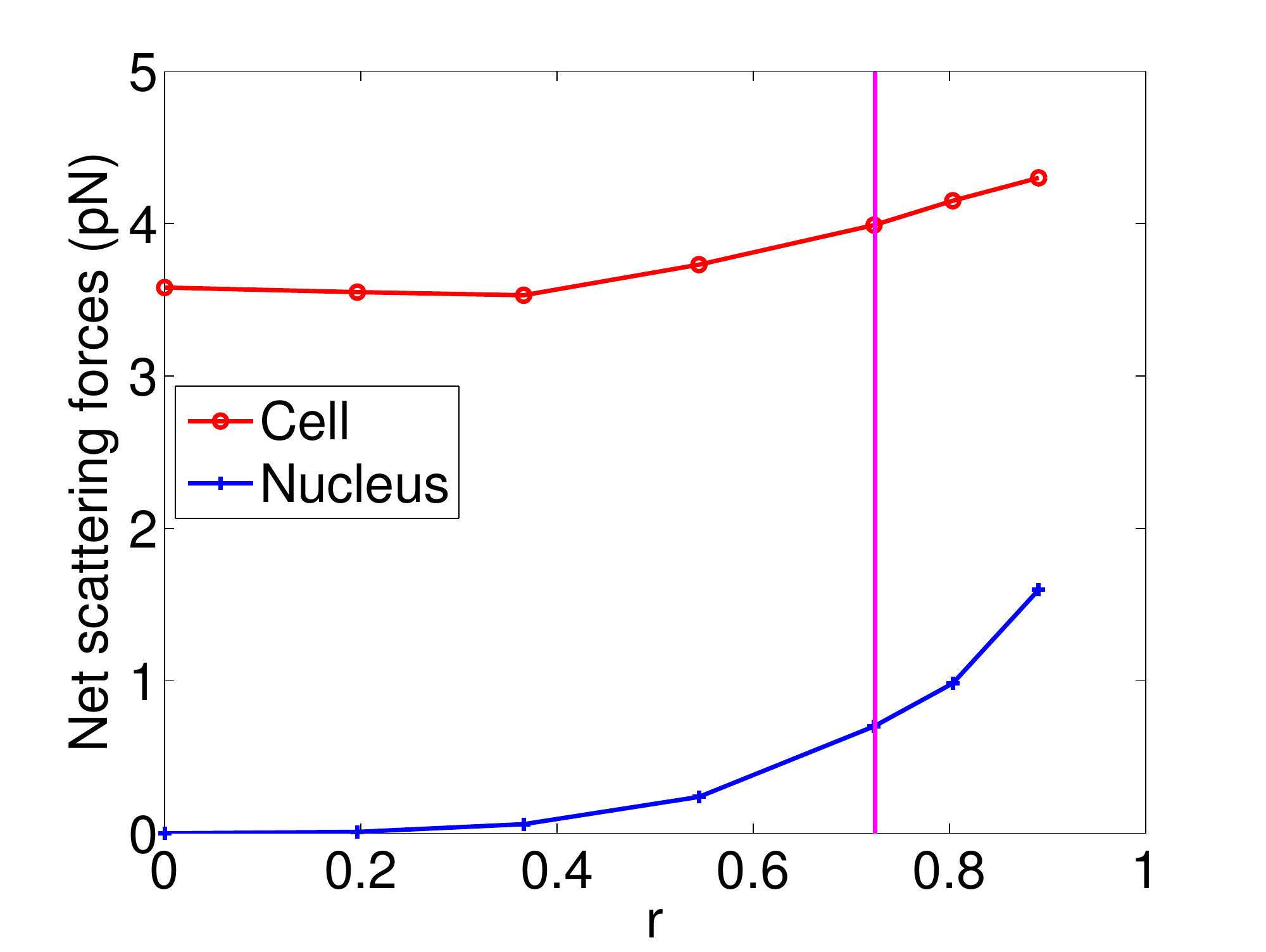} &
\includegraphics[width=0.45\textwidth]{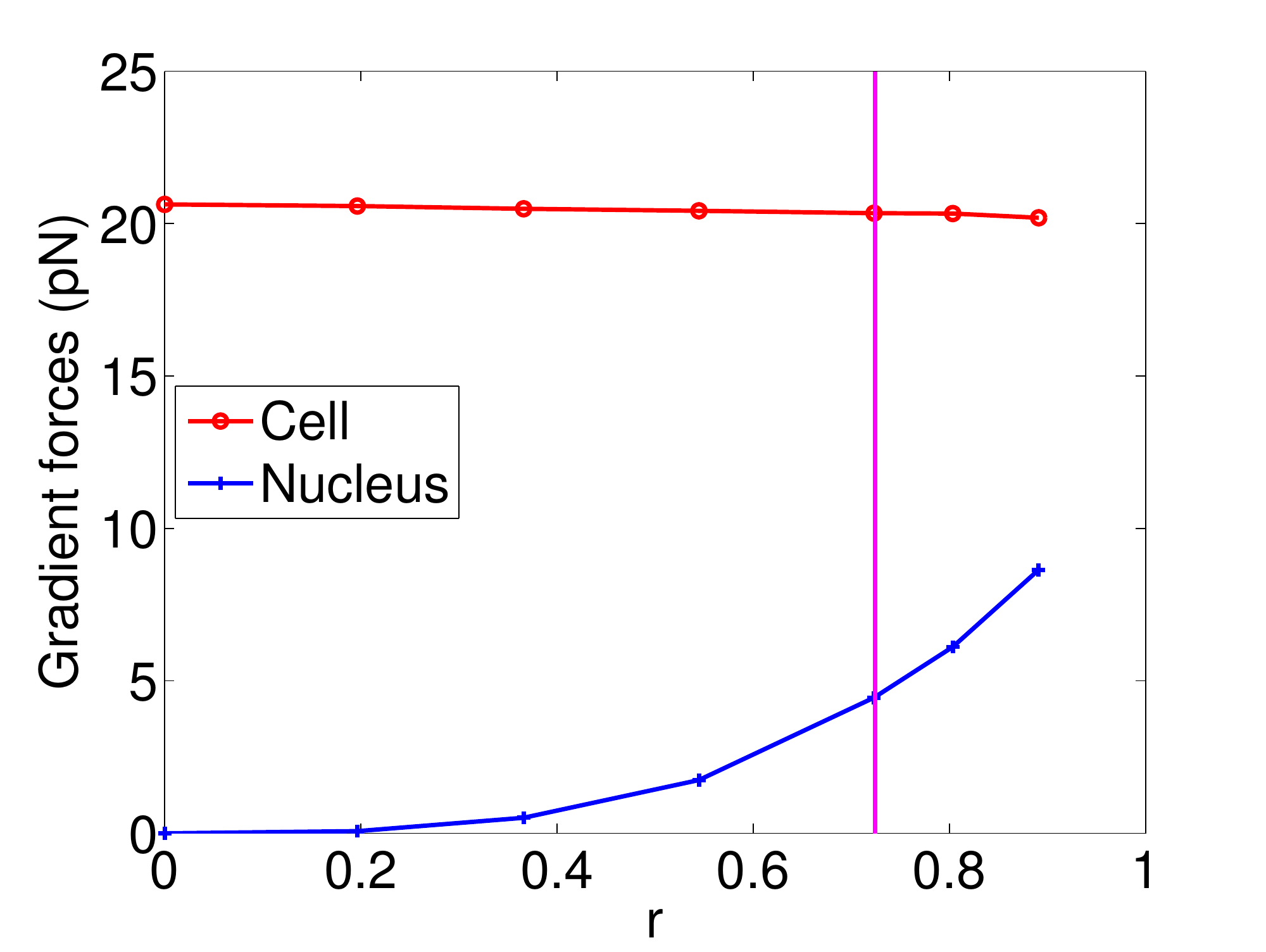} 
\end{tabular}

\caption{Comparison of the magnitude of optical forces on a CHO cell for different nucleus radii: (left) net scattering forces (right) gradient forces. The vertical line in magenta corresponds to the nominal CHO cell radius ratio
of $r = 0.72$.} 
\label{fig:fnet} 
\end{figure}

The variation of the net scattering and gradient forces on the cell and its nucleus
is shown in Fig.~\ref{fig:fnet} where we clearly see the effect of the size 
of the nucleus on the total net force induced on the cell surface. As the nucleus size increases,
the cell will be exposed to larger forces for the same laser power. The 
nucleus acts as a lens that focuses the light rays and leads to higher 
optical forces at the nucleus and cell back surfaces. We finally
note that we can determine the net scattering and gradient forces
on a CHO cell of nominal size from the curves in Fig.~\ref{fig:fnet}
using a vertical line (shown in magenta) that
corresponds to the nominal CHO cell radius ratio
of $r = 0.72$.

\subsection{Influence of nucleus size on net cell deformation}

From our calculations, it is clear that the presence of 
a nucleus has a significant impact on the initial 
optical force distribution as these forces deform and stretch 
CHO cells. Changes in cell shape lead to a new force distribution.
To calculate these, optical forces 
are added as body forces to the surrounding fluid. DRT is then employed
to update the optical force distribution as the cell shape is changing
until steady state when the elastic and applied optical forces 
are equal. 

\begin{figure}[htbp]
\centering 
\begin{tabular}{clclcl}
\includegraphics[width=0.40\textwidth]{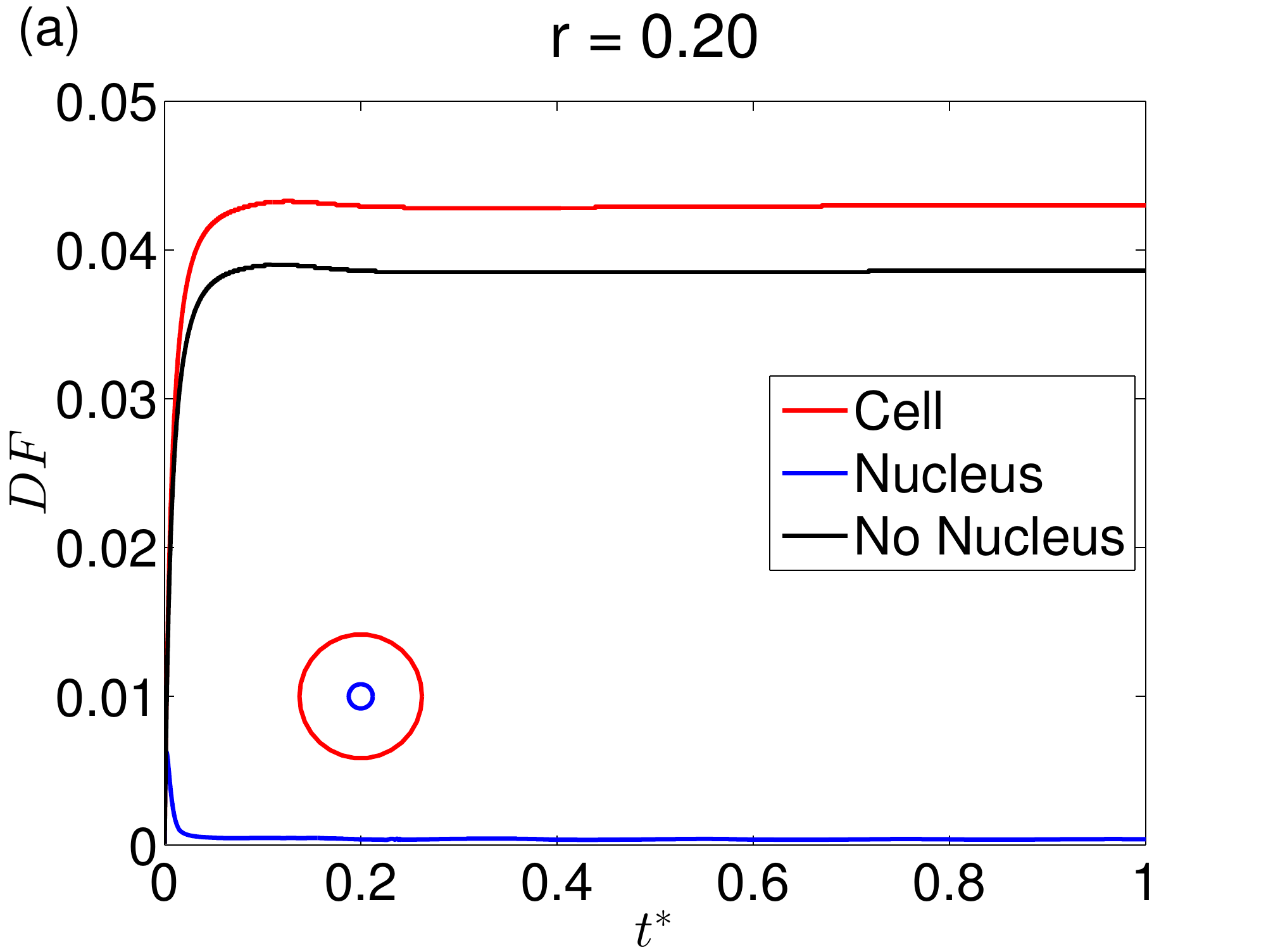} &&
\includegraphics[width=0.40\textwidth]{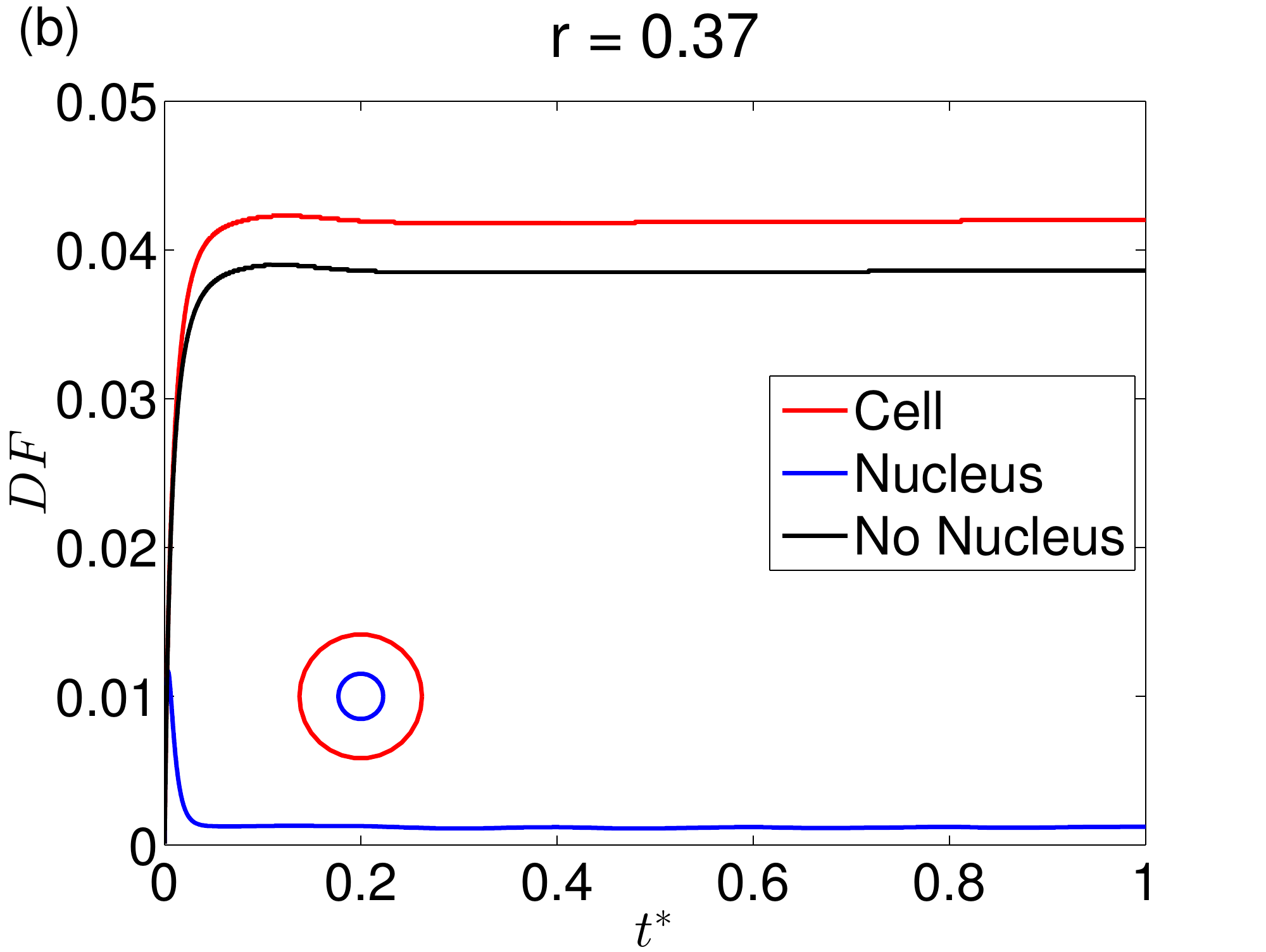} \\
\includegraphics[width=0.40\textwidth]{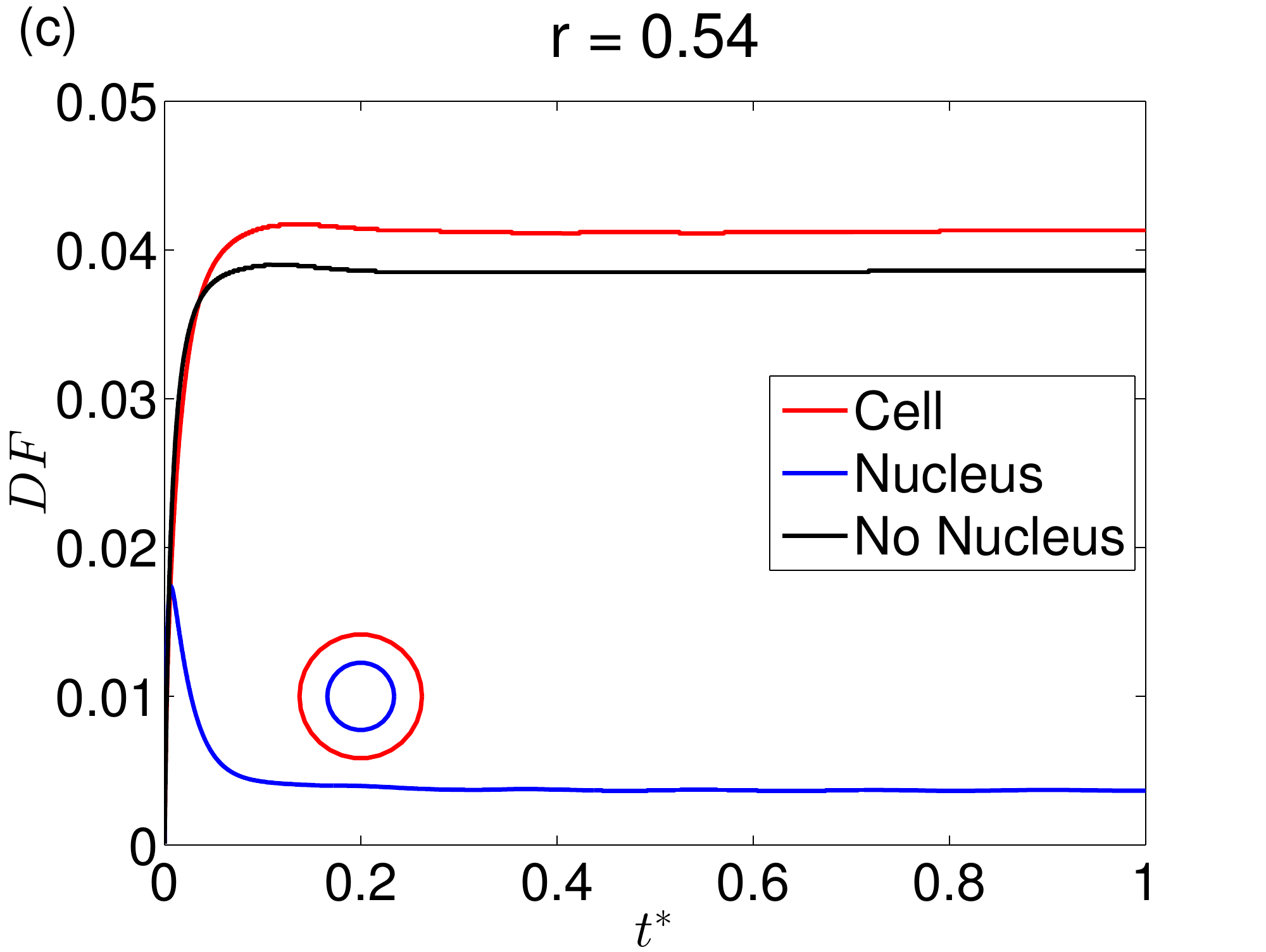} &&
\includegraphics[width=0.40\textwidth]{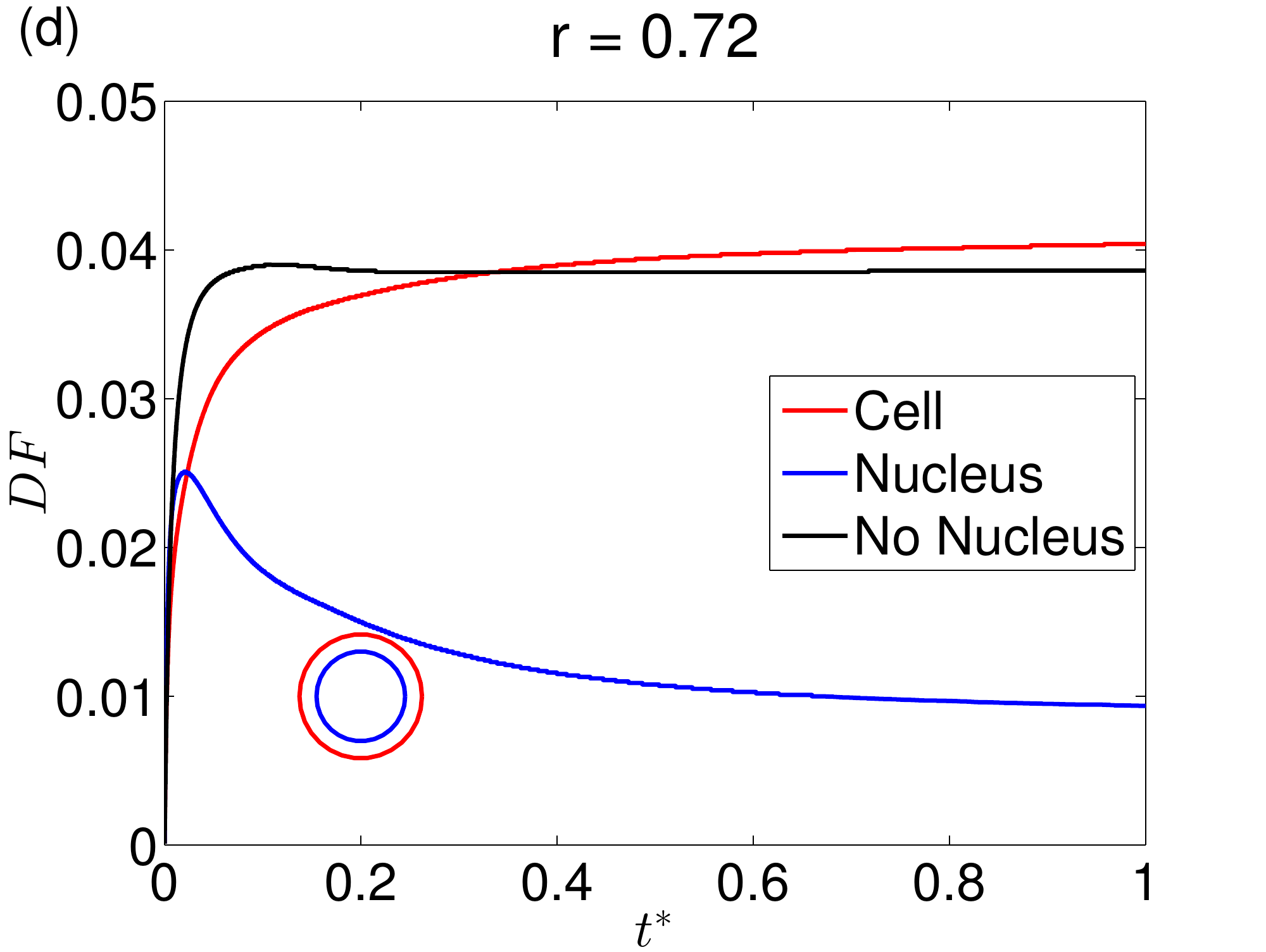} \\
\includegraphics[width=0.40\textwidth]{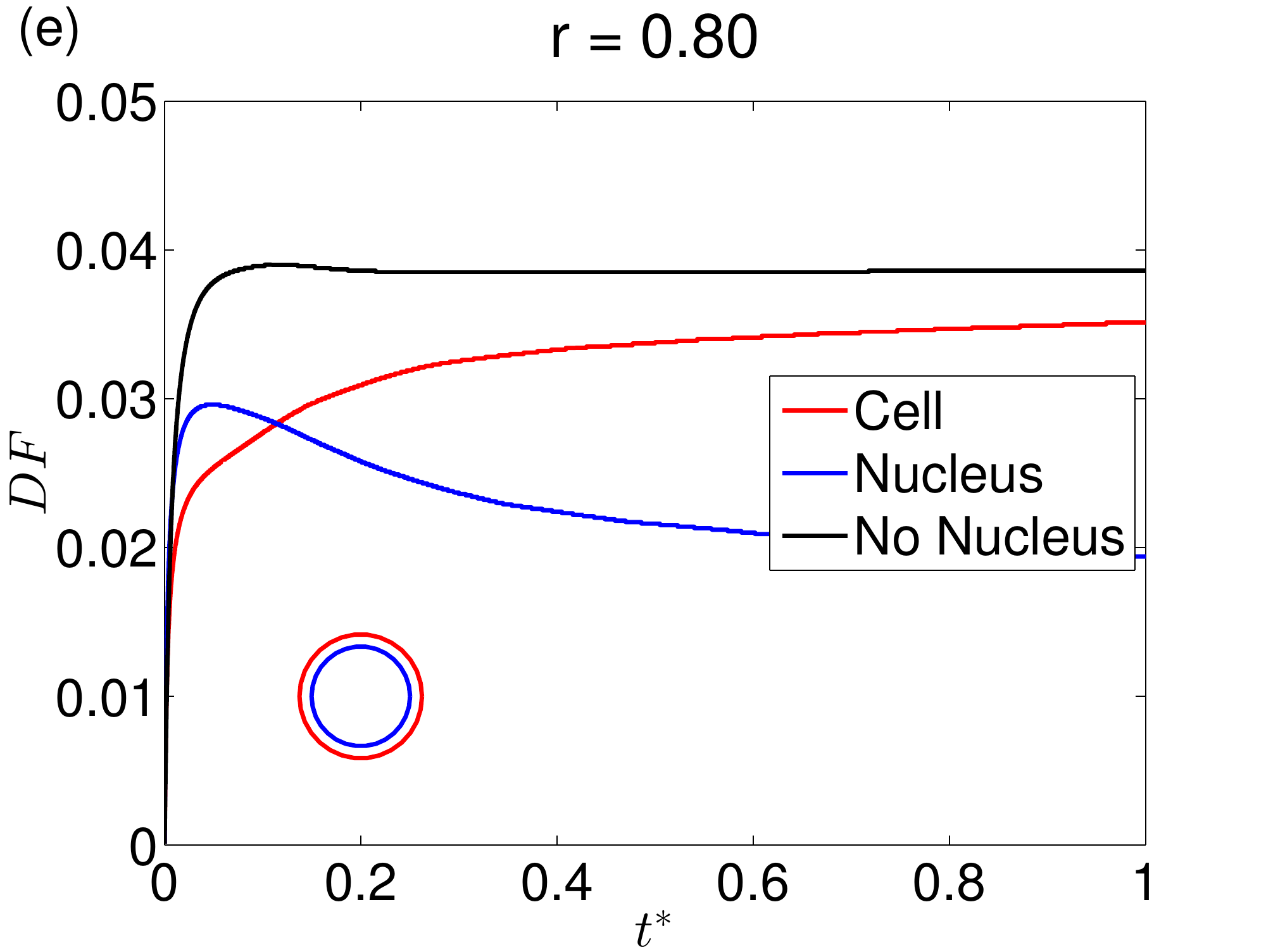} &&
\includegraphics[width=0.40\textwidth]{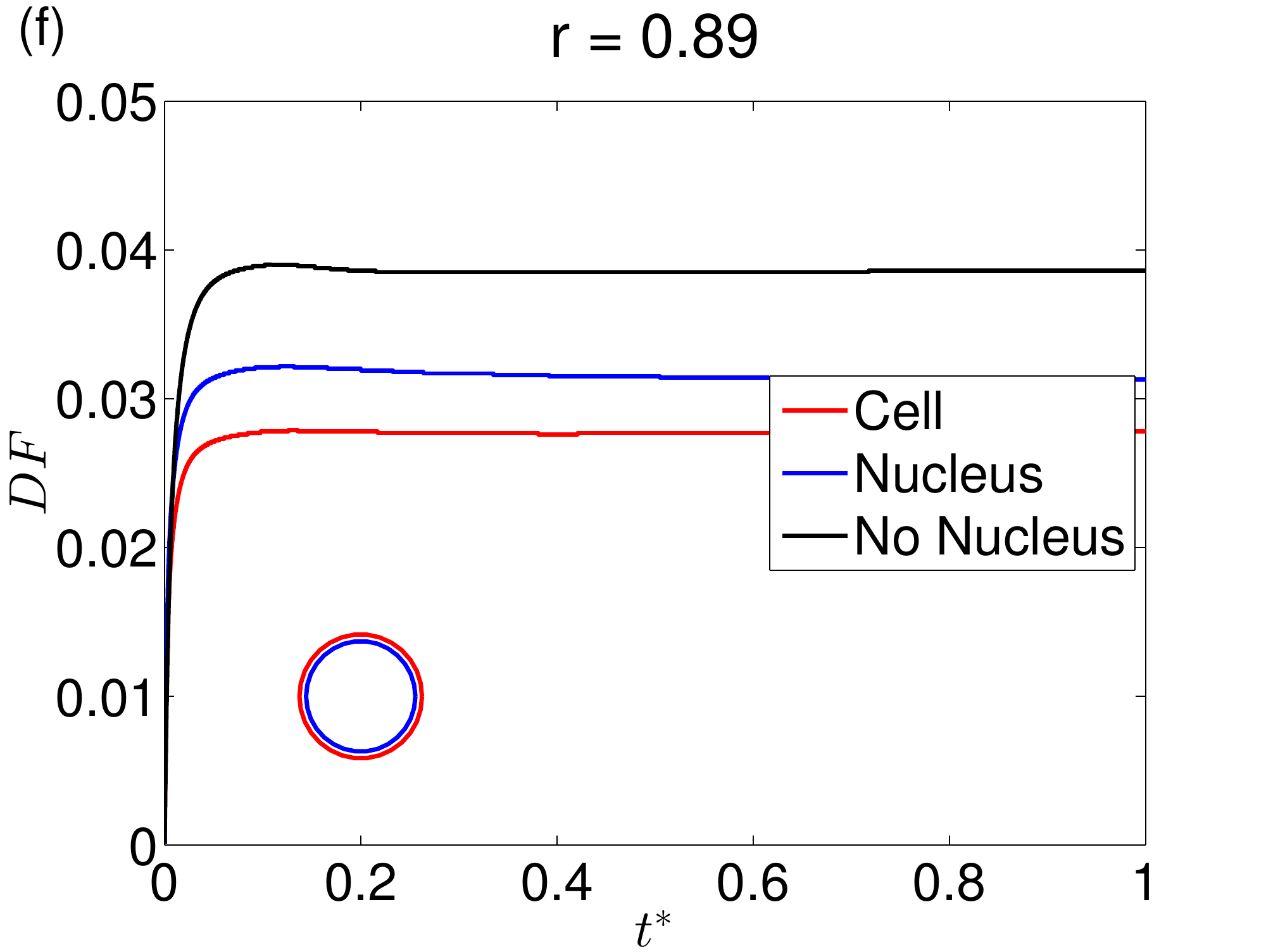} \\
\end{tabular}
\caption{Evolution of net deformation $DF$ (of both a CHO cell and its nucleus) for different nucleus size as indicated. Evolution of net deformation of cell with no-nucleus is
also shown for reference.} 
\label{fig:dft} 
\end{figure}

The net deformation of both the cell and the nucleus is quantified using the Taylor 
parameter deformation $DF$ shown in Fig.~\ref{fig:dft}. 
The figure indicates that in the case of small nuclei, $DF$ of the cell increases to a steady value 
that is higher than the reference case of no-nucleus (shown on all panels for comparison).
This is due to the slight decrease in gradient forces that lead to more
deformation in the z-direction and thus higher net deformation.
$DF$ of the nucleus, however, is negligible due to the small magnitudes of the optical forces created.

For larger nuclei, $DF$ shows similar trends where the cell deforms
until a steady state shape is reached but with net deformation lower than  
the nucleus-free case. In this, we see that the net deformation of the cell decreases 
and the net deformation of the nucleus increases with $r$.

\begin{figure}[htbp]
\centering 
\includegraphics[width=0.55\textwidth]{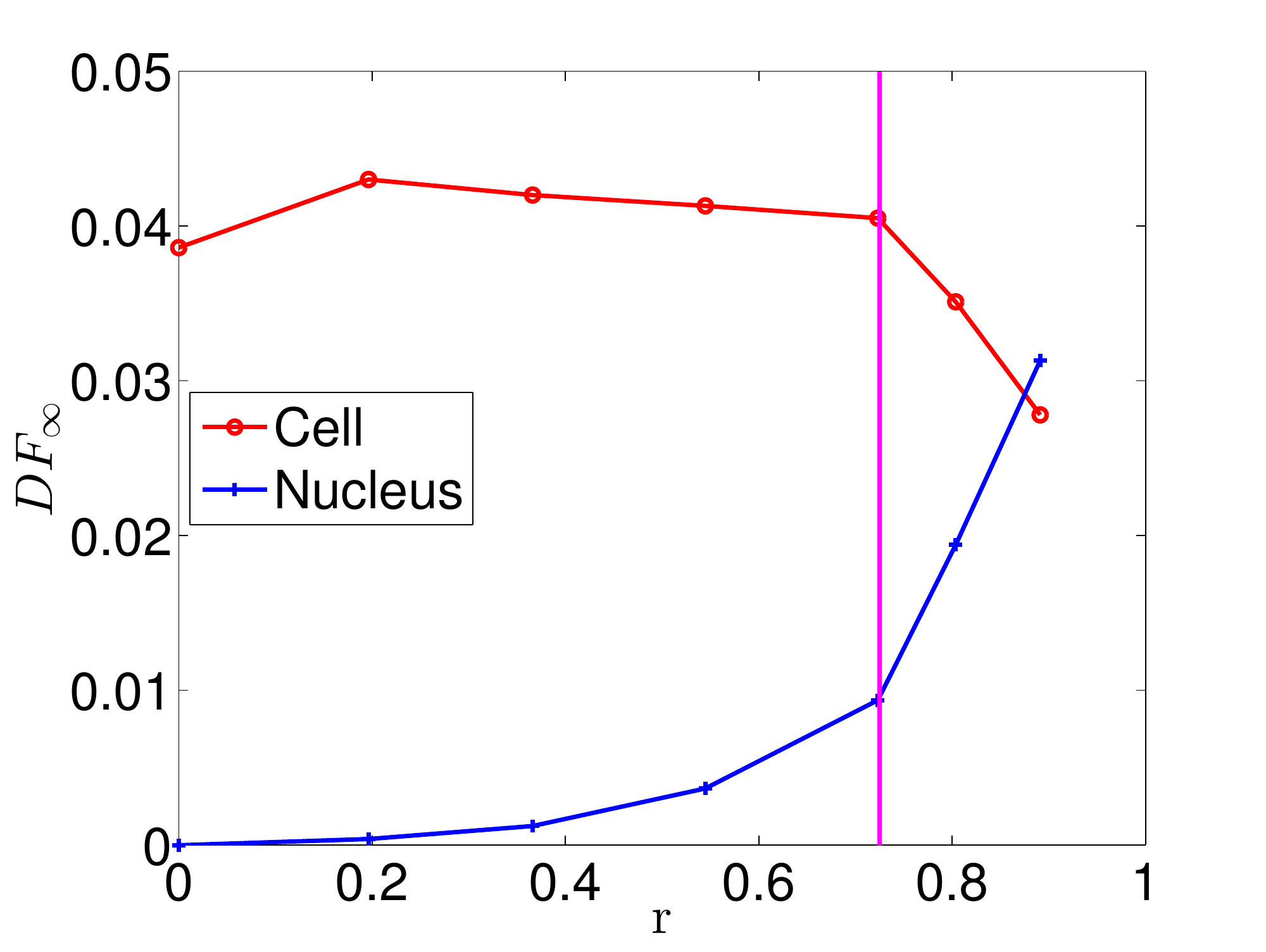}
\caption{Comparison of net deformation of a CHO cell and its nucleus for different nucleus radii.
The vertical line in magenta corresponds to the nominal CHO cell radius ratio
of $r = 0.72$.
} 
\label{fig:dfnet} 
\end{figure}

To summarize the results discussed above, we show in Fig.~\ref{fig:dfnet} 
the steady state net deformation $DF_{\infty}$ of both the cell and the 
nucleus. 
Here, we clearly see that
$DF_{\infty}$ of the cell is initially larger than the 
$DF_{\infty}$ of the nucleus-free case but then decreases as the radius 
ratio increases. At the same time, $DF_{\infty}$ of the nucleus
increases with the radius ratio. The two curves eventually 
intersect when the radius of the nucleus becomes comparable to the radius
of the cell.  We also
note here that we can determine the steady state net deformation $DF_{\infty}$
of a CHO cell of nominal size from the curves in Fig.~\ref{fig:dfnet}
using a vertical line (shown in magenta) that
corresponds to the nominal CHO cell radius ratio
of $r = 0.72$.

$DF$ calculations for CHO cells show a clear relationship
between the size of the nucleus and the steady state net cell and nuclues
deformation. 
As the size of the nucleus increases, the steady state deformation
increases due to the increase in the scattering forces applied at the cell
surface. It is therefore expected that the corresponding
deformation increases as well. Fig.~\ref{fig:dfnet} shows also that as the 
size of the nucleus increases, the steady state deformation of the cell 
decreases.  Moreover, as the size of the nucleus approaches the size of the cell, 
the relative deformation of the nucleus, as characterized by the Taylor deformation parameter surpasses that of the cell. 
The relationship between the size of the nucleus 
and the deformation of the cell means that cells with larger 
nuclei show less deformation when optically stretched than cells with 
smaller nuclei. Another observation is that the cell with no nucleus deformed 
to a steady state of $DF$ value that is smaller than the value for the smallest 
radius ratio but larger than the value for the largest radius ratio.

\section{Conclusions}
Chinese hamster ovary cell is of interest to the biomedical community 
because of their use in recombinant protein therapeutics. 
Unfortunately, no experiments have been performed to stretch this line of cells
in optical traps. Few experiments, however, were performed to describe deformation of cell 
nucleus. In these experiments researchers have studied the interaction of cells with 
topographically patterned material surfaces to show
the changes in shape, function, and viability of the cells. Only few of
these studies, however, indicated possible impact on the behavior of organelles.
For instance, Dalby \emph{et al.} \cite{Dalby2003} quantified cell and
nuclear morphology with light and fluorescence microscopy and showed a slight elongation of 
the nuclei in grooves. Yamauchi \emph{et al.} also observed the deformation of cancerous cells/nucleus
and their migration in capillaries of mince \cite{Yamauchi2005}. This study aimed
to find the minimum diameter of capillaries where cancer cells are able to migrate.
They measured the diameter of both the cell and its nucleus for this purpose.

In our work, we modeled the deformation of
Chinese hamster ovary cells by single 
diode-laser bar optical stretchers. For this purpose,
we extended the recently developed Dynamic Ray-Tracing method
to determine the stress distribution induced by the applied optical forces
on cells that have a nucleus. Our results showed that the presence of a nucleus has a 
major effect on the force distribution on the cell surface and 
the net deformation.  We also showed and quantified the effect of nucleus size on 
the net applied force as well as on cell deformation.
We are working effectively on setting up experiments to stretch CHO cells
and compare our numerical data with experimental results.
\section{Acknowledgments}
The authors would like to acknowledge financial support provided by the National Institute of Health grant R01 AI079347-04.
This work used the Extreme Science and Engineering Discovery Environment (XSEDE), which is supported by National Science Foundation grant number OCI-1053575.

\bibliographystyle{osajnl}
%

%

\end{document}